\documentclass[preprint,12pt]{elsarticle}

\usepackage{amssymb}
\usepackage{amsmath}
\usepackage{subcaption}
\usepackage{multirow}
\usepackage{etoolbox}
\usepackage{lineno}
\journal{***}

\begin{document}

\begin{frontmatter}

%% Title, authors and addresses

%% use the tnoteref command within \title for footnotes;
%% use the tnotetext command for theassociated footnote;
%% use the fnref command within \author or \affiliation for footnotes;
%% use the fntext command for theassociated footnote;
%% use the corref command within \author for corresponding author footnotes;
%% use the cortext command for theassociated footnote;
%% use the ead command for the email address,
%% and the form \ead[url] for the home page:
%% \title{Title\tnoteref{label1}}
%% \tnotetext[label1]{}
%% \author{Name\corref{cor1}\fnref{label2}}
%% \ead{email address}
%% \ead[url]{home page}
%% \fntext[label2]{}
%% \cortext[cor1]{}
%% \affiliation{organization={},
%%             addressline={},
%%             city={},
%%             postcode={},
%%             state={},
%%             country={}}
%% \fntext[label3]{}

\title{Position reconstruction and surface background model for the PandaX-4T detector}

%% use optional labels to link authors explicitly to addresses:
%% \author[label1,label2]{}
%% \affiliation[label1]{organization={},
%%             addressline={},
%%             city={},
%%             postcode={},
%%             state={},
%%             country={}}
%%
%% \affiliation[label2]{organization={},
%%             addressline={},
%%             city={},
%%             postcode={},
%%             state={},
%%             country={}}

%% Author affiliation
% \affiliation{organization={},%Department and Organization
%             addressline={}, 
%             city={},
%             postcode={}, 
%             state={},
%             country={}}
\author[a]{Zhicheng Qian}
\author[a]{Linhui Gu}
\author[b]{Chen Cheng}
\author[a]{Zihao Bo}
\author[a]{Wei Chen}
\author[a,c,d,e]{Xun Chen}
\author[e,f]{Yunhua Chen}
\author[g]{Zhaokan Cheng}
\author[c]{Xiangyi Cui}
\author[h]{Yingjie Fan}
% ...

\affiliation[a]{School of Physics and Astronomy, Shanghai Jiao Tong University, Key Laboratory for Particle Astrophysics and Cosmology (MoE), Shanghai Key Laboratory for Particle Physics and Cosmology, Shanghai 200240, China}
\affiliation[b]{School of Physics, Beihang University, Beijing 102206, China}
\affiliation[c]{New Cornerstone Science Laboratory, Tsung-Dao Lee Institute, Shanghai Jiao Tong University, Shanghai 201210, China}
\affiliation[d]{Shanghai Jiao Tong University Sichuan Research Institute, Chengdu 610213, China}
\affiliation[e]{Jinping Deep Underground Frontier Science and Dark Matter Key Laboratory of Sichuan Province}
\affiliation[f]{Yalong River Hydropower Development Company, Ltd., 288 Shuanglin Road, Chengdu 610051, China}
\affiliation[g]{Sino-French Institute of Nuclear Engineering and Technology, Sun Yat-Sen University, Zhuhai, 519082, China}
\affiliation[h]{Department of Physics, Yantai University, Yantai 264005, China}
% ...

% \collaboration{PandaXy-4T Collaboration}%
%% Abstract
\begin{abstract}
We report the position reconstruction methods and surface background model for the PandaX-4T dark matter direct search experiment. This work develops two position reconstruction algorithms: template matching (TM) method and photon acceptance function (PAF) method. Both methods determine the horizontal position of events based on the light pattern of secondary scintillation collected by the light sensors. After a comprehensive evaluation of resolution, uniformity, and robustness, the PAF method was selected for position reconstruction, while the TM method was employed for verification. The PAF method achieves a bulk event resolution of 1.0 mm and a surface event resolution of 4.4 mm for a typical $S2$ signal with a bottom charge of 1500 PE (about 14 keV). The uniformity is around 20\%. Robustness studies reveal average deviations of 5.1 mm and 8.8 mm for the commissioning run (Run0) and the first science run (Run1), respectively, due to the deactivation of certain PMTs. A data-driven surface background model is developed based on the PAF method. The surface background is estimated to be $0.09 \pm 0.06$ events for Run0 (0.54 tonne$\cdot$year) and $0.17 \pm 0.11$ events for Run1 (1.00 tonne$\cdot$year).
\end{abstract}

%%Graphical abstract
% \begin{graphicalabstract}
% %\includegraphics{grabs}
% \end{graphicalabstract}

%%Research highlights
% \begin{highlights}
% \item Research highlight 1
% \item Research highlight 2
% \end{highlights}

%% Keywords
\begin{keyword}
%% keywords here, in the form: keyword \sep keyword
Dual-phase time projection chamber \sep Xenon detector \sep Reconstruction \sep Surface background \sep Dark matter
%% PACS codes here, in the form: \PACS code \sep code

%% MSC codes here, in the form: \MSC code \sep code
%% or \MSC[2008] code \sep code (2000 is the default)

\end{keyword}

\end{frontmatter}

%% Add \usepackage{lineno} before \begin{document} and uncomment 
%% following line to enable line numbers
%% \linenumbers

%% main text
%%

%% Use \section commands to start a section
\section{Introduction}
\label{sec:intro}
%引言，本底分析的重要性
The existence of dark matter (DM) is strongly supported by cosmological and astronomical observations \cite{Bertone2005}, yet its nature remains elusive.
The Weakly Interacting Massive Particle (WIMP) has become one of the most competitive candidates for dark matter, as suggested by numerous theories beyond the Standard Model of particle physics \cite{APPECreport}.
Dark matter direct detection experiments, typically situated deep underground, are designed to be particularly sensitive to dark matter within a mass range from approximately GeV/$c^{2}$ to 100 TeV/$c^{2}$.
% , via the nuclear recoil (NR) of the target nucleus
In recent years, large-scale liquid xenon time projection chambers (TPCs) have led the way in enhancing detection sensitivity~\cite{Meng2021,pandax2024,Aprile2023_nTFirst,Aalbers2023_LZFirst}. 
Recently, the PandaX-4T experiment has reported the dark matter search results of the commissioning run (Run0) and the first science run (Run1)~\cite{pandax2024}.
This work focuses on the position reconstruction and surface background model in PandaX-4T WIMP analysis.

In this work, two horizontal position reconstruction algorithms are developed for the PandaX-4T combined analysis using data of both Run0 and Run1~\cite{Meng2021,pandax2024}: the template matching (TM) method and the photon acceptance function (PAF) method. Both methods aim to determine the horizontal scattering position of each event based on the information collected by the top PMT array. 
The TM method uses a set of templates of light patterns, which are generated by optical simulation, to match the measured light distribution pattern. The PAF method utilizes a model that describes the relation between the light distribution pattern and the scattering position, enabling it to calculate the position based on the measured light distribution. 
A more accurate position reconstruction enables the development of a reliable surface background model.
Surface background is primarily caused by radioactive isotopes on the surface of PTFE reflectors, such as $\rm{}^{210}Pb$, $\rm{}^{210}Bi$ and $\rm{}^{210}Po$. In the energy region of interest (ROI) for WIMP analysis, $\rm{}^{210}Pb$  is the dominant contributor. In WIMP analysis, determining the fiducial volume (FV) is critical for suppressing background noise and improving detection sensitivity. To optimize the FV, especially at the side boundary, it is essential to obtain a comprehensive understanding of the surface background event distribution. In this work, a data-driven surface background model is established by utilizing the characteristic peak from the $\rm{}^{210}Po$ alpha decay, which reflects the $S2$ features of surface events.

In this paper, the content is organized as follows. 
Sec.~\ref{sec_p4detector} illustrates the geometry of the PandaX-4T detector.
Sec.~\ref{sec_Algorithm} introduces position reconstruction algorithms, including optical simulation, TM and PAF construction methods, and advanced corrections applied in PandaX-4T analysis.
Sec.~\ref{sec_Performance} describes the performance of the reconstruction algorithms via various properties, including bulk and surface event resolutions, uniformity, and robustness.
Sec.~\ref{sec_surface_model} presents a data-driven surface background model based on the previously introduced position reconstruction method and provides the final number of surface background events in PandaX-4T analysis.

\section{The PandaX-4T detector}
\label{sec_p4detector}
\begin{figure}[t]
    \centering
    \begin{subfigure}{0.45\textwidth}
        \centering
        \includegraphics[height=6cm]{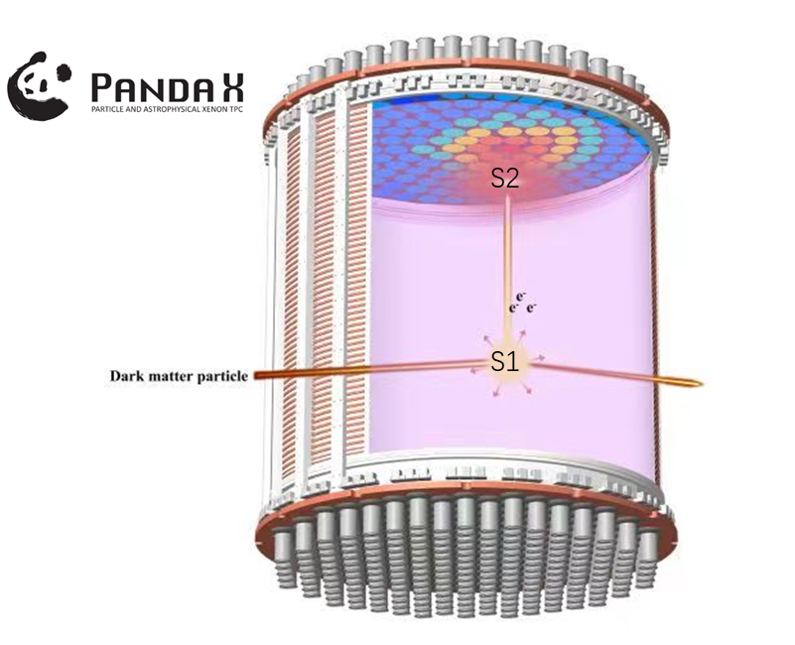}
        \caption{3D model of the detector structure.}
        \label{fig:TPC}
    \end{subfigure}
    \hfill
    \begin{subfigure}{0.45\textwidth}
        \centering
        \includegraphics[height=6cm]{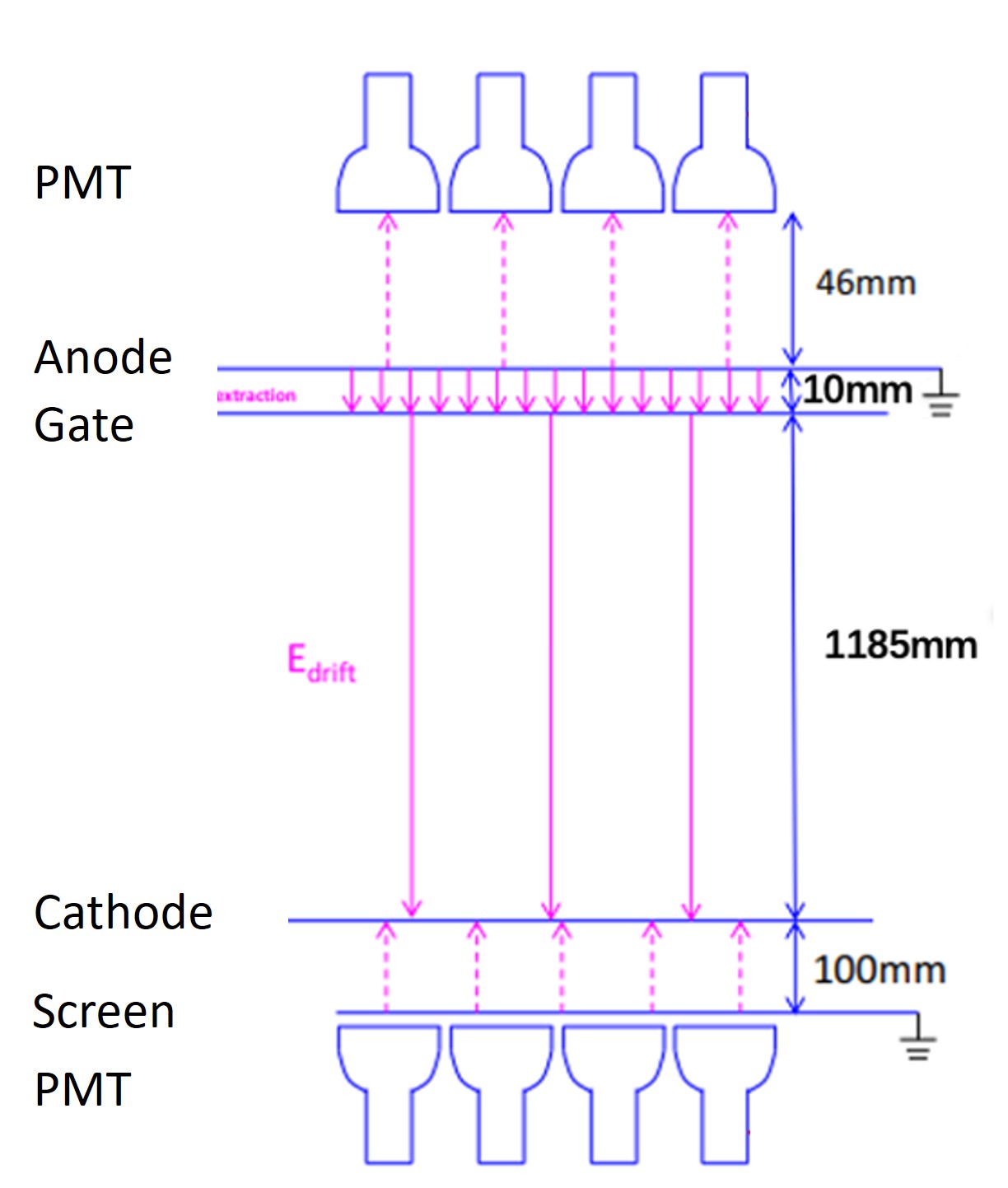}
        \caption{2D schematic diagram of the detector layout.}
    \label{fig:layout}
    \end{subfigure}
    \caption{Overview of the PandaX-4T detector design.}
    \label{fig:p4detector}
\end{figure}

%介绍padnax-4t tpc 运作原理
The PandaX-4T detector, with a sensitive target of 3.7 tonnes of liquid xenon, is located in the B2 hall of the newly expanded CJPL-II~\cite{LI2015576,Kang2010}. 
Schematic diagrams of PandaX-4T detector are shown in figure~\ref{fig:p4detector}. The central cylindrical
TPC has a transverse diameter of 1.2 m and a height of 1.3 m. As shown in figure~\ref{fig:p4detector}, the TPC is enclosed by highly reflective Polytetrafluoroethylene (PTFE) reflector panels, with 24 wall panels on the side and two pannels on the top and bottom. Inside TPC, highly transparent electrodes are designed to generate electric fields. From bottom to top, these include screen mesh, cathode grid, gate mesh, and anode mesh, with separations of 1185 mm and 10 mm, respectively (figure~\ref{fig:layout}). The gas-liquid interface is between the anode and the gate. Additionally, approximately 60 copper shaping rings, consisting of copper strips arranged in a circular configuration outside the PTFE wall panels as shown in figure~\ref{fig:TPC}, surround the TPC. These rings are designed to maintain a uniform vertical drift field within the detector.
Two arrays of Hamamatsu R11410-23 three-inch photomultipliers (PMTs) are located at the top and bottom of the TPC to collect light signals. The PMTs are fixed onto the copper plates outside the Teflon reflector panels at both ends.

When a WIMP collides with a xenon atom, it generates a prompt scintillation light signal ($S1$) and ionizes electrons. These electrons, driven by the vertical electric field, drift upwards, undergo electroluminescence amplification, and produce a secondary signal ($S2$) that retains the horizontal position of the initial scattering event.
These $S1$ and $S2$ signals are collected by the PMT arrays.
The scattering vertex in three-dimensions can then be reconstructed based on the collected photon distribution pattern (light pattern) on the top PMT array and the time difference between the $S1$ and $S2$ signals.
The reconstruction of the three-dimensional spatial position information of scattering events plays a crucial role in event identification, background suppression, and energy reconstruction.

\begin{figure}[t]
        \centering
        \includegraphics[width=\linewidth]{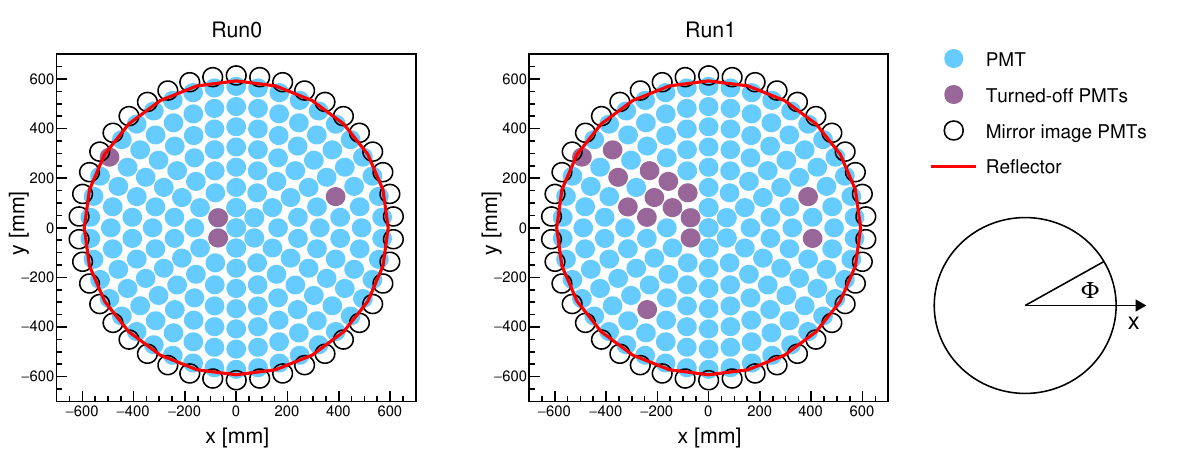}
        \caption{Operation status of top PMT arrays for Run0 (left) and Run1 (right). Different markers indicate the status of the PMTs: blue circles represent operational PMTs, purple circles denote turned-off PMTs, and hollow circles represent image PMTs constructed in the PAF method. The corresponding polar coordinates are shown in the bottom-right corner.}
        \label{fig_RMT_map} 
    % \end{minipage}
\end{figure}

Currently, the PandaX-4T experiment includes Run0 data with an exposure of 0.54 tonne$\cdot$year and Run1 data with an exposure of 1.00 tonne$\cdot$year~\cite{pandax2024}. Figure~\ref{fig_RMT_map} shows the distribution and operation status of the top PMT array during the PandaX-4T data-taking process. There are 169 top PMTs placed in a concentric circular pattern. For clarification, the origin of the Cartesian coordinate system is defined at the center of the TPC, indicating the $z$-positions of the gate and cathode are 592.5 mm and -592.5 mm, respectively. 
For the polar coordinate system, the azimuthal angle ($\Phi$) is defined with the positive $x$-axis as the zero point.

\section{Position reconstruction algorithms}
\label{sec_Algorithm}
In addition to the conventional reconstruction method known as center of gravity (COG) , PandaX-4T used TM and PAF methods to enhance position reconstruction performance.

\subsection{Template matching method}
\label{subsec_TM}
The TM method is developed based on optical simulation~\cite{PR_DEAP-3600}. The simulated light patterns of top PMT array are used to reconstruct the horizontal position of each signal. To ensure reconstruction accuracy, it is crucial that the simulated light patterns closely match those obtained from actual data. The optical simulation for PandaX-4T is carried out using BambooMC~\cite{BambooMC2021}, a GEANT4-based Monte Carlo (MC) package. The geometry of the PandaX-4T detector described in Sec.~\ref{sec_p4detector} is established in BambooMC. 

To more accurately simulate the behavior of the detector, especially the light patterns of the PMTs, tuning to certain parameters of the optical simulation is required, including the reflectivity and refractive index of the PTFE reflectors, etc. Root mean square (RMS) can serve as a metric to quantify the spreading in a pattern's distribution, which is defined as
\begin{equation}\label{RMS}
% RMS =\sqrt{\frac{\sum_{i}^{}[(X_{pmt,i}-X_{pos})^{2}+(Y_{pmt,i}-Y_{pos})^{2}]\cdot Q_{pmt,i}}{\sum_{i}^{}Q_{pmt,i}}},
\mathrm{RMS} =\sqrt{\frac{\sum_{i}|\Vec{r}_{i}-\Vec{r}|^{2}\cdot Q_{i}}{\sum_{i}Q_{i}}},
\end{equation}
where $Q_{i}$ is the charge received by the $i^{\rm th}$ PMT, $\Vec{r}_{i}$ is the position of the $i^{\rm th}$ PMT, and $\Vec{r}$ is the position of certain event. In simulation, $\Vec{r}$ represents the location where the photon is generated, whereas in the data, $\Vec{r}$ denotes the reconstructed position of an event.

% In the simulation, the light sources are placed in the gas phase between the liquid surface and the anode, where S2 events usually occur. The horizontal distribution of photons takes into account the transverse diffusion of the electrons during their drifting and is modeled as a 2D Gaussian with a standard deviation of \textcolor{red}{$\sigma=3$ mm. (reference?)}

\begin{figure}[h]
    \centering
    \begin{minipage}[t]{\linewidth}
        \centering
        \includegraphics[width=0.48\linewidth]{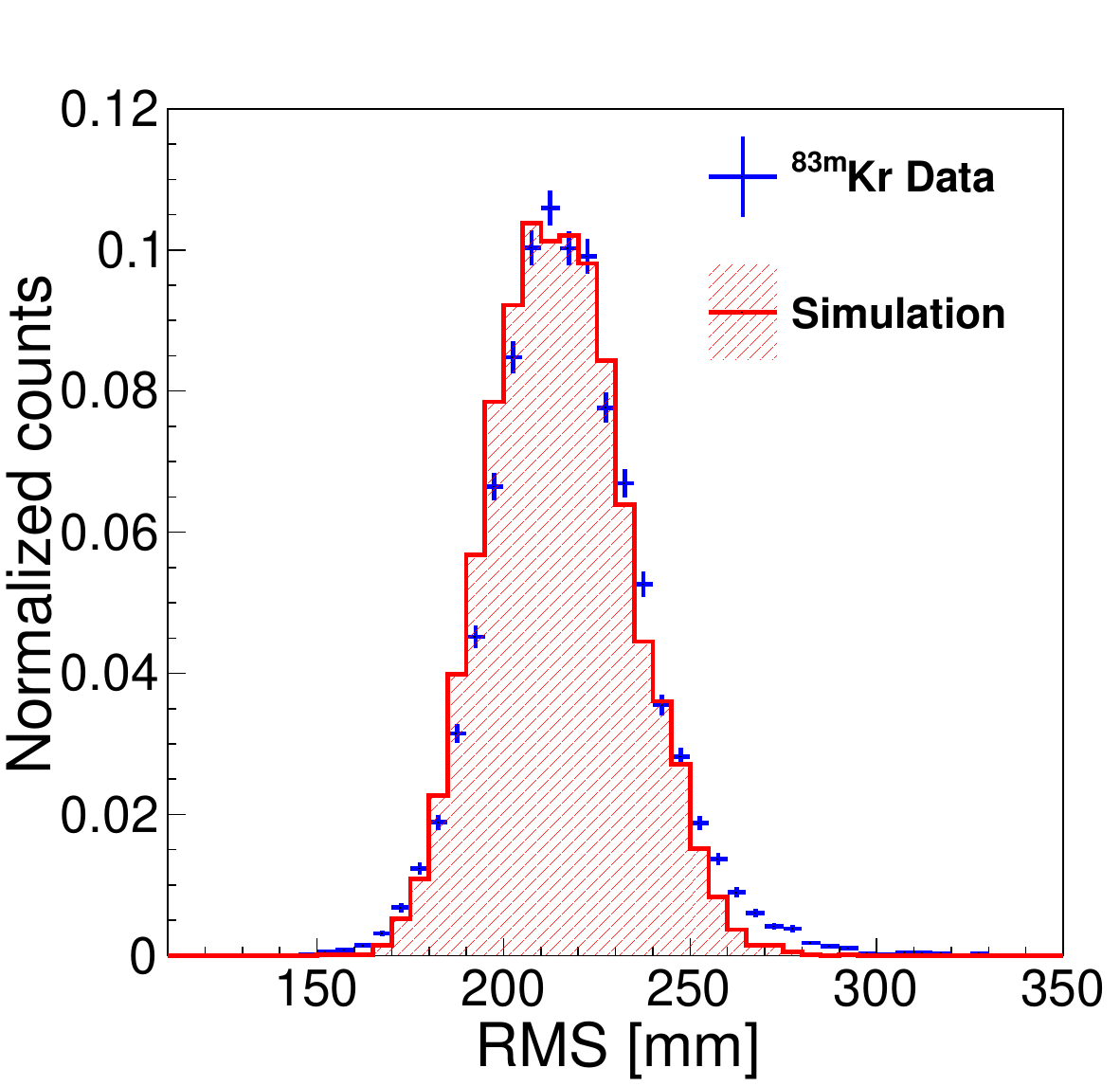}
        \caption{RMS comparison between data and simulation.}
        \label{fig_RMS} 
    \end{minipage}
\end{figure}

$^{83m}$Kr calibration events with 41.5 keV energy peak can be used as a sample for refining simulation parameters. Using the $S2$ distribution of $^{83m}$Kr events as input, the simulation parameters are adjusted to match the RMS distributions between the simulation and the data. Figure~\ref{fig_RMS} shows the RMS comparison, with the mean values for the data and the simulation being 217$\pm$21 and 215$\pm$18, respectively. The difference of approximately 1\% indicates good consistency. The reflectivity and refractive indexes of PTFE are set to be 99\% and 1.61 separately. Other optical parameters are summarized in table~\ref{tab_pars}.

\begin{table}[htbp] 
\small
\centering
  \begin{tabular}{ccccc}
    \hline
    Part&Absorption length & Refractivity &Rayleigh scattering length\\
    \hline
    Gas xenon     &100.0 m      &1.00 &100.0 m\\
    Liquid xenon  &10.0 m       &1.61 &50.0 cm\\
    Anode         &0.1255 cm    &1.00 & /     \\
    Gate \& screen          &0.134 cm     &1.61 & /     \\
    Cathode       &0.35 cm      &1.61 & /     \\
    \hline
  \end{tabular}
  \caption{Optical properties of major materials involved in simulation.}
  \label{tab_pars}
\end{table}

Based on optical simulations, about $10^{4}$ templates are generated. Each template represents the expected distribution of light on the top PMT array with a given event position.
The horizontal position of each signal is reconstructed by matching its $S2$ light pattern against the generated templates.
The maximum likelihood (ML) estimation is used to assess how well the template matches the real light pattern. The position of the template with maximum likelihood most closely approximates the actual scattering point of an event. 
The likelihood function used in the TM method is expressed as
\begin{equation}
\label{L}
L\left(\vec{r}_{t}\right)=\prod_{i=1}^{m}p\left ( n_{i},p_{i}\right )=\prod_{i=1}^{m}\frac{e^{-Np_{i}}\left ( Np_{i} \right )^{n_{i}}}{n_{i}!} ,
\end{equation}
where $p_{i}$ is the probability of a photoelectron (PE) being detected by the $i^{\rm th}$ PMT for the template generated at position $\vec{r}_{t}$, $n_{i}$ is the number of PEs detected by the $i^{\rm th}$ PMT, $N = \sum\limits_{i=1}^{m} n_{i}$ is the total number of PEs and $m$ is the number of all operational top PMTs.  
In order to efficiently find the best-matched template, the stochastic gradient descent algorithm~\cite{gd-algorithms} instead of iterating over all templates is applied. A 10 mm spacing between template positions is selected to strike a balance between computation time and reconstruction resolution.

The final reconstructed position is obtained by performing a weighted average of the template with the maximum likelihood and its surrounding templates:
% \begin{equation}\label{Xsmear}
% \vec{r_{reco}}=\frac{\sum_{j}^{}w_{j}\cdot \vec{r_{t, j}}}{\sum_{j}^{}w_{j}},  \rm{and}
% \end{equation}

\begin{equation}\label{Xsmear}
\vec{r}_{\text{reco}} = \frac{\displaystyle\sum_{j} w_{j} \cdot \vec{r}_{t, j}}{\displaystyle\sum_{j} w_{j}}, \quad
w_{j}=\frac{{\rm ln}L_{j}}{{\rm ln}(L_{j}/L_{\rm max})\cdot |\Vec{r}_{j}-\Vec{r}_{\rm max}|} ,
\end{equation}
where $L_{j}$ and $\Vec{r}_{j}$ are the likelihood value and position of a surrounding template $j$, while $L_{\rm max}$ and $\Vec{r}_{\rm max}$ are that of the best-matched template. The surrounding templates are within a radius of 15 mm around the best-matched template.

\subsection{Photon acceptance function method}
The PAF method describes the fraction of light that is collected by each PMT as a function of the event's position~\cite{PAF-ZEPLIN, MLofPos, Akerib_2018, LINDOTE2007200, osti_1212164, PhysRevD.100.052014}. This concept has been used to develop algorithms for position reconstruction in scintillation detectors~\cite{MLofPos}.
The light fraction of each PMT is calibrated by optical simulations mentioned in Sec.~\ref{subsec_TM}.
% These simulations provide the advantage of accurately known sample event positions.

%% from Dan's note
The single-variable PAF used in this work is similar to that used in PandaX-II~\cite{2021PAF}. The basic analytical form $\eta_{i}^{0}$ of the PAF is given by~\cite{PAF-ZEPLIN}
\begin{equation}\label{PAF_initial}
\eta_{i}^{0}\left(\iota _{i}\right)=A_{i}\cdot exp\left ( -\frac{a_{i}\cdot \frac{\iota _{i}}{r_{i}}}{1+\left (\frac{\iota _{i}}{r_{i}}\right )^{1-\alpha _{i}}} -\frac{b_{i}}{1+\left (\frac{\iota _{i}}{r_{i}}\right )^{\alpha _{i}}}\right ),
\end{equation}
where $i$ is the PMT index, and $\iota _{i}$ is the distance from the scattering point to the center of the $i^{\rm th}$ PMT. The other parameters, including $A_{i}$, $\alpha_{i}$, $r_{i}$, $a_{i}$, $b_{i}$, are fitting parameters. 
% The PAF trained for a single PMT is shown in figure \ref{PAF_function}.

In order to correct the asymmetry caused by PTFE reflection at the border, the concept of  ``mirror image PMTs'' (figure~\ref{fig_RMT_map}) is introduced. $\eta_{i}^{0}$ is separated into two parts, the reduced object $\eta_{i}$, and the corresponding image, $\eta_{i, im}$. For PandaX-4T analysis, the PAF is extended to
\begin{equation}\label{PAF_correct}
\eta_{i}\left(\iota _{i}\right)=\frac{1}{1+\omega }\cdot \left ( 1-\rho   \right )\cdot\eta_{i}^{0}\left(\iota _{i}\right), \rm{and}
\end{equation}
\begin{equation}\label{PAF_im}
\eta_{i,im}\left(\iota _{i,im}\right)=\omega \cdot\eta_{i}\left(\iota _{i,im}\right),
\end{equation}
where the factor $\frac{1}{1+\omega}$ represents the proportion of photons received by the PMT that are attributed to the mirrored PMT. $\omega$ is non-trivial for PMTs near the boundary and zero for the inner PMTs. The term $\rho $ serves as a correction factor for the global reflection effects. The minus sign before $\rho $ in Eq.~\ref{PAF_correct} represents a cut-off correction.
The statistical inference of the position is also done by ML estimation as deduced in the literature~\cite{MLofPos}. The modified likelihood function considering image PMTs is expressed as
\begin{equation}\label{PAF_likelihood_withimagePMT}
\begin{aligned}
\ln L\left ( \vec{r} \right )&=\sum_{\rm edge}^{}\left ( \frac{1}{1+\omega } \cdot q_{i}\cdot \ln \frac{\eta _{i}\left ( \iota _{i} \right )}{P\left ( \vec{r} \right )}+ \frac{\omega}{1+\omega } \cdot q_{i}\cdot \ln \frac{\eta _{i,im}\left ( \iota _{i,im} \right )}{P\left ( \vec{r} \right )} \right ) 
\\ & + \sum_{\rm inner}^{}q_{i}\cdot \ln  \frac{\eta _{i}\left ( \iota _{i} \right )}{P\left ( \vec{r} \right )} ,
\end{aligned}
\end{equation}
where $q_{i}$ is the collected $S2$ charge by the $i^{\rm th}$ PMT, and the probability $P\left(\vec{r}\right) = \sum_{i}(\eta_{i}+\eta_{i,im})$. The position with the maximum likelihood, obtained by scanning $(x, y)$ for each event with a step size of 0.01 mm, is taken as the reconstructed position of the event.

\subsection{Advanced corrections}
\label{subsec_cor}
\begin{figure}[htbp]
\centering
    \begin{minipage}[t]{0.38\linewidth}
        \centering
        \includegraphics[width=\linewidth]{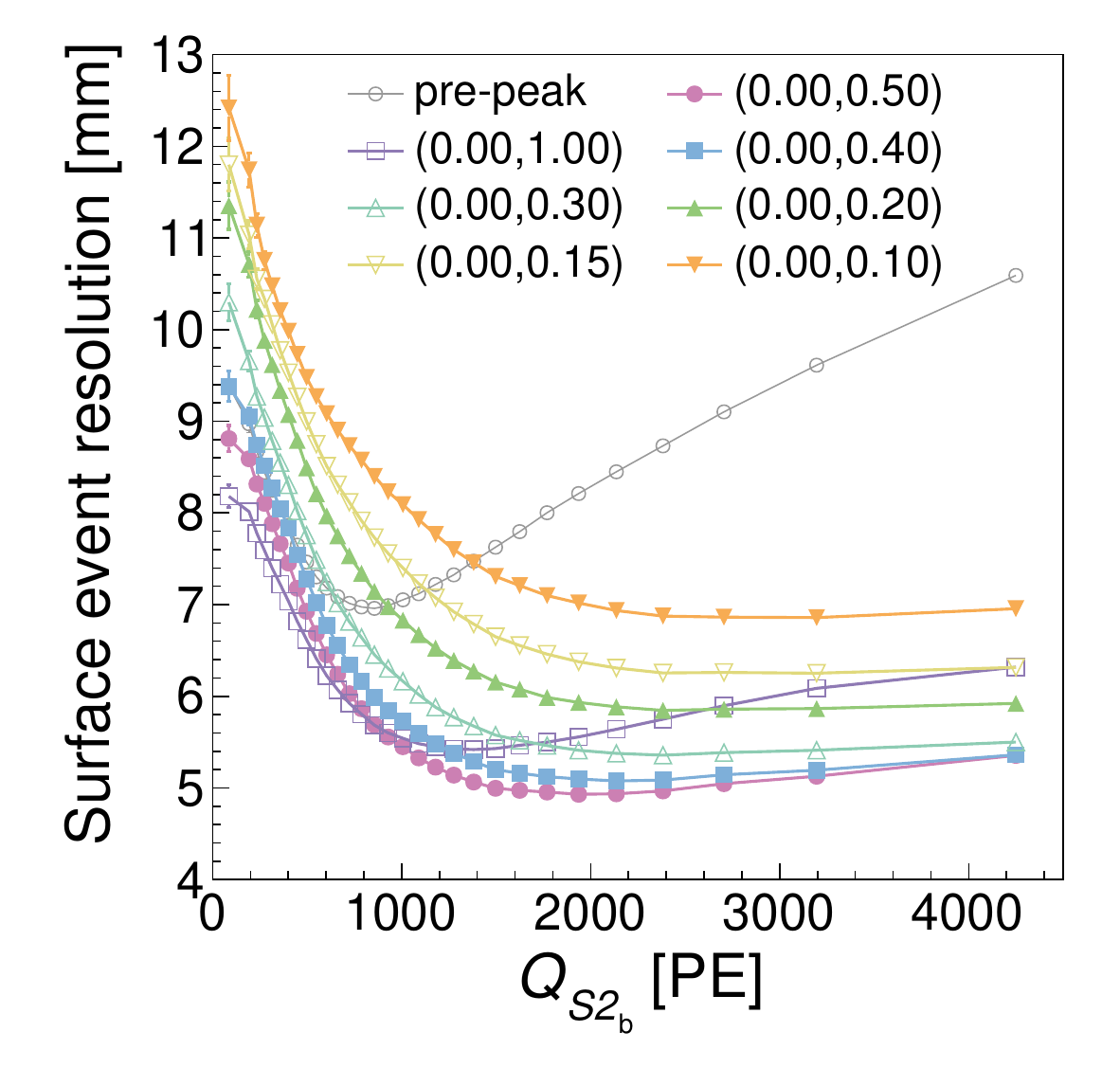}
        \caption{Surface event resolutions for different CDF ranges.}
        \label{fig_cdf}
    \end{minipage}
    \hfill
    \begin{minipage}[t]{0.57\linewidth}
        \centering
        \includegraphics[width=\linewidth]{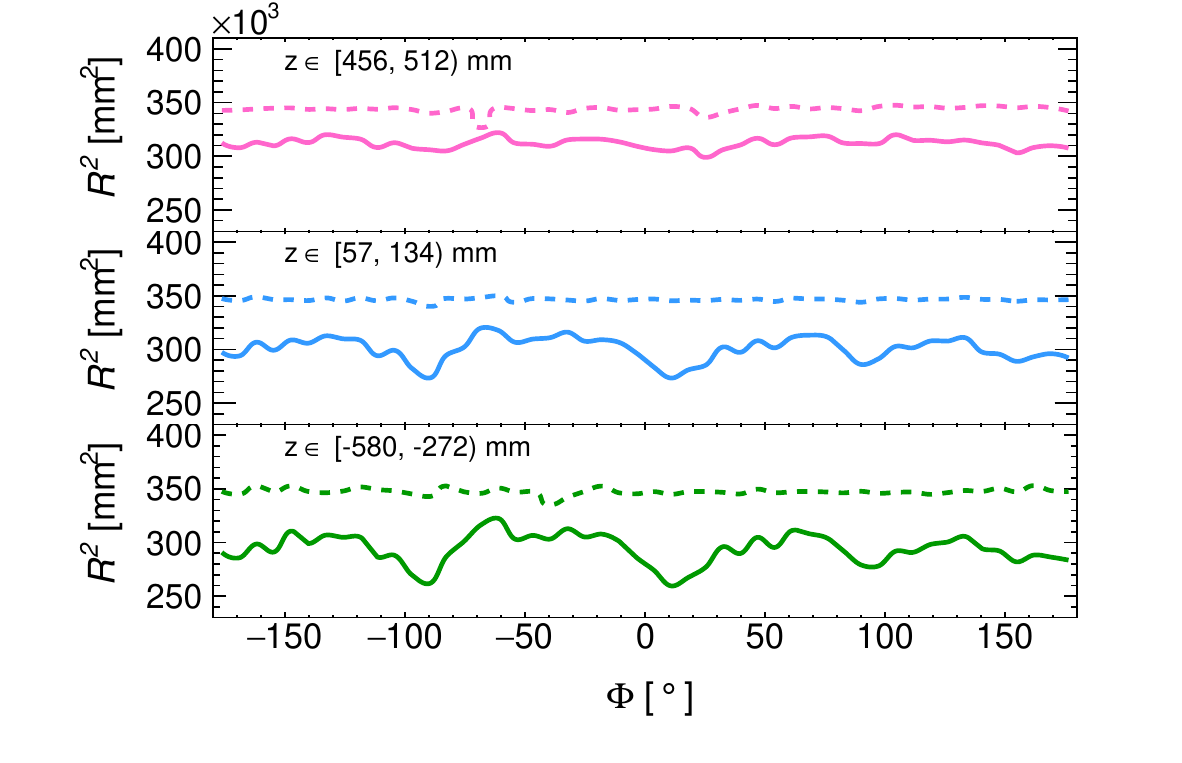}
        \caption{Azimuthal angle distribution of surface events at different vertical positions. (solid lines: before GPC; dashed lines: after GPC)}
        \label{fig_stretch}
    \end{minipage}
\end{figure}

Two advanced corrections are applied to the algorithms: partial waveform reconstruction (PWR) and geometric position correction (GPC). The PWR is designed to address light pattern dispersion caused by the tailing of $S2$ signals, which leads to deviations in reconstructed positions. 
% These tails may consist of delayed electrons and additional noise charges that do not originate from the $S2$ signals themselves. 
These tails may consist of additional noise charges that do not originate from the $S2$ signals themselves. 
% The delayed electrons are dragged out as electron clusters pass through the gas-liquid interface. 
% Possible sources of noise charges include secondary electron signals generated by the photoelectric effect on electrodes and PMT afterpulses.
Possible sources of noise charges include secondary electron signals generated by photoionization and PMT afterpulses.
For each signal, instead of the total waveform, PWR reconstructs the position using a waveform segment, selected based on the cumulative distribution function (CDF).
Given the signal tailing, the segment's endpoint is set where CDF $\leq$ 50\%, meaning that the first half of the waveform contributes most to the reconstruction. To quantify the effect of different CDF ranges, surface event resolution is measured using the width of the $^{210}$Po characteristic Gaussian peak. Figure~\ref{fig_cdf} illustrates the performance across various CDF ranges. For comparison, a ``pre-peak" curve is shown, which integrates the waveform from the starting point up to its peak.
Optimal performance is achieved when the CDF range is set between 0\% and 50\%. Notably, the surface event resolution is evaluated as a function of $S2$ bottom charge ($Q_{S2_{\rm b}}$), since $S2$ top charge ($Q_{S2_{\rm t}}$) may underestimate the true charge due to saturation effects in the top PMTs, making $Q_{S2_{\rm b}}$ a more reliable choice. The ratio of the top charge to the bottom charge is approximately 4:1 when the PMTs are not saturated.

In both Run0 and Run1, a phenomenon is observed where events at the edge exhibit a noticeable deviation of the reconstructed radial position $R$ towards the center of TPC at $15^\circ$ and $-90^\circ$ (figure~\ref{fig_stretch}). This inward shift becomes more pronounced for events closer to the bottom of the TPC. This may due to the non-uniformity of the local drift electric field yet no obvious deformations of the reflectors were observed. To more accurately approximate the true positions, GPC is performed. $^{210}$Po alpha events, primarily orginating from the PTFE surface, are used as samples to eliminate $\Phi$ dependence. Additionally, $^{83m}$Kr events, which are expected to be uniformly distributed in TPC, are used as samples to align the reconstructed positions with the geometric positions. Figure~\ref{fig_stretch} demonstrates the effects of GPC at different vertical positions.

\section{Position reconstruction performance}
\label{sec_Performance}
In this section, the performance of different position reconstruction algorithms will be examined in terms of resolutions, uniformity and robustness. 

\subsection{Resolutions}
\label{subsec_resolutions}
\begin{figure}[h]
\centering
    \begin{minipage}[t]{0.49\linewidth}
        \centering
        \includegraphics[width=\linewidth]{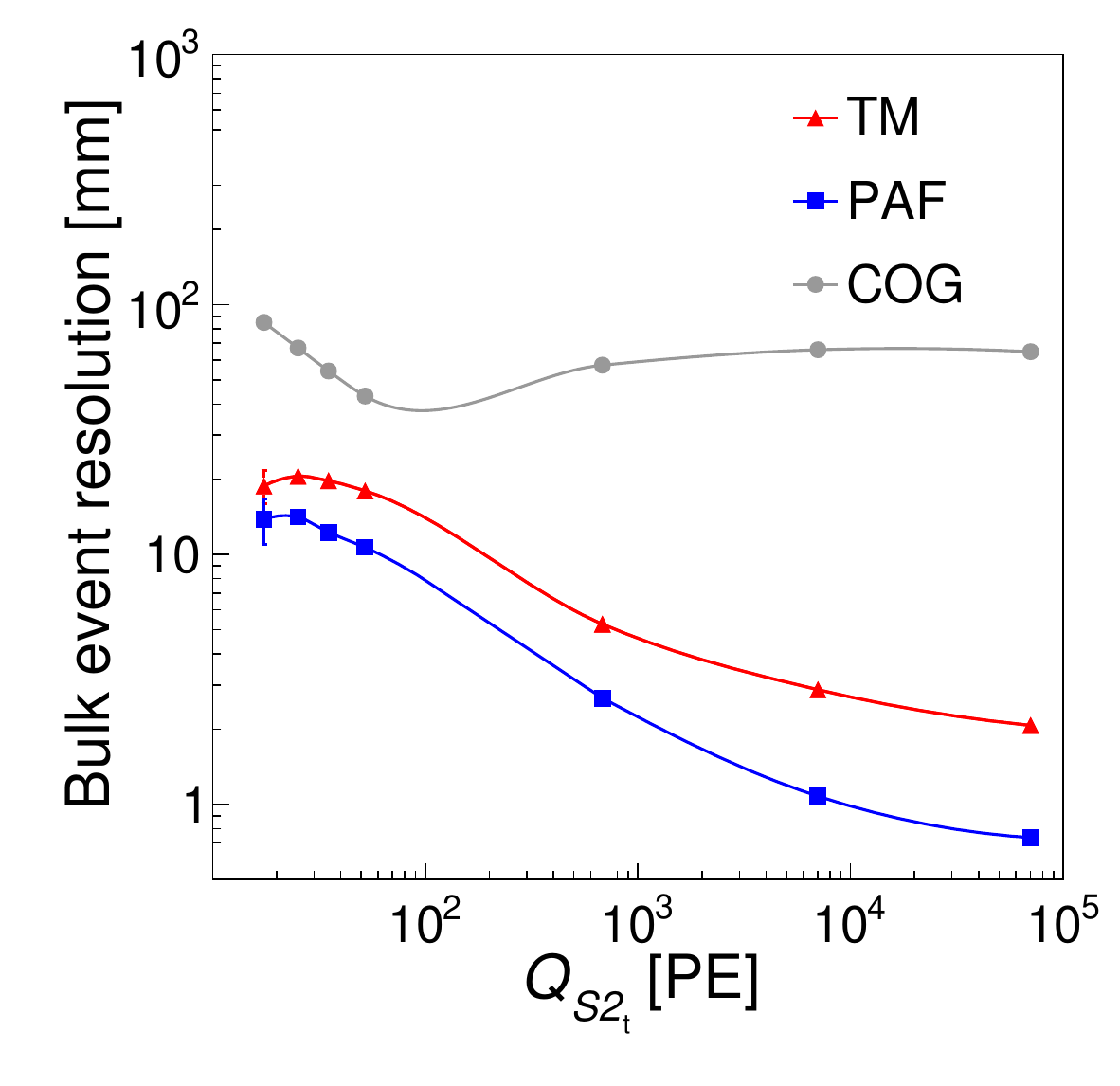}
        \caption{Bulk event resolution on $S2$ top charge ($Q_{S2_{\rm t}}$) for different reconstruction methods.}
        \label{fig_bulk_res}
    \end{minipage}
    \hfill
    \begin{minipage}[t]{0.49\linewidth}
        \centering
        \includegraphics[width=\linewidth]{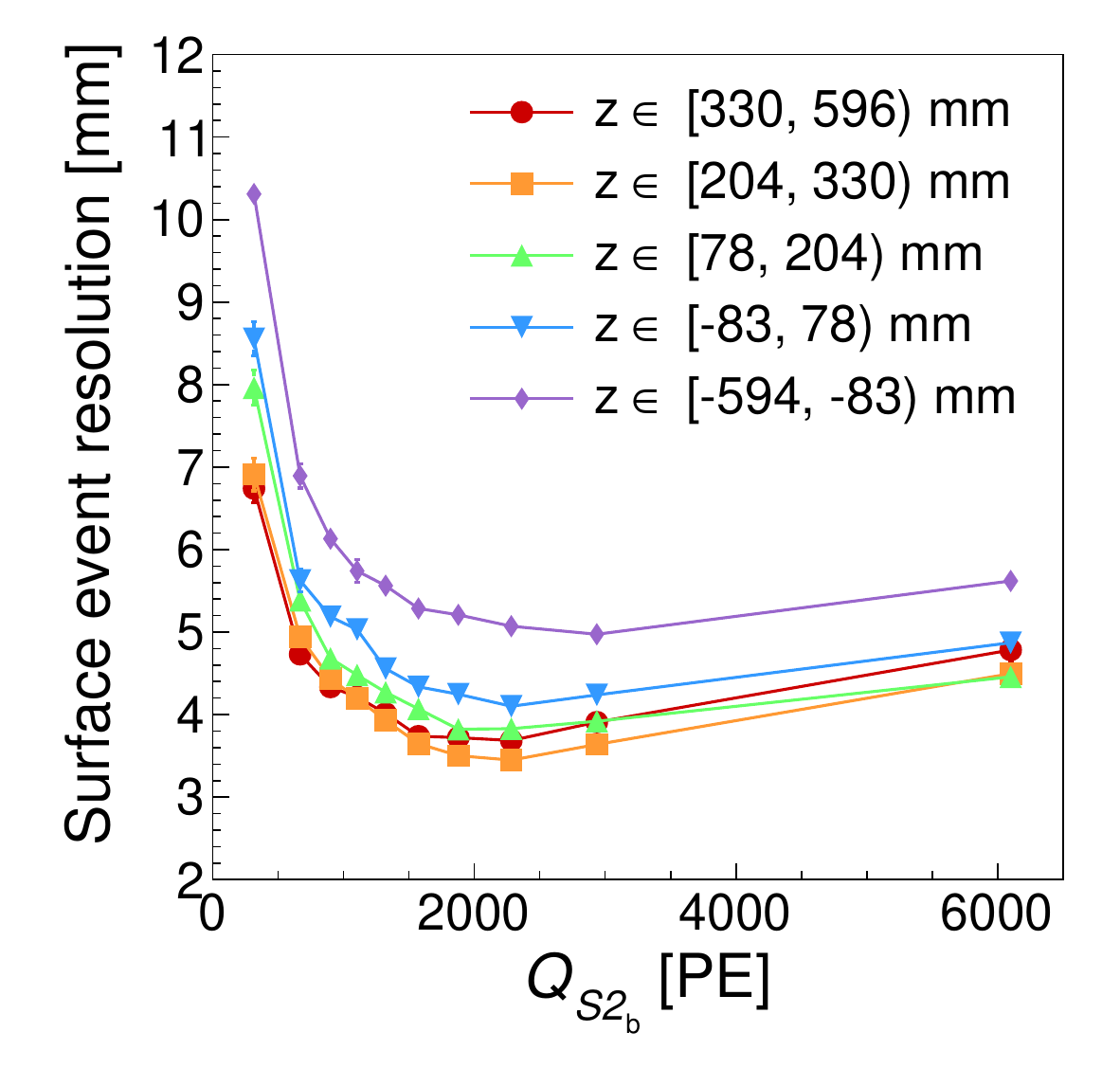}
        \caption{Surface event resolution on $S2$ bottom charge ($Q_{S2_{\rm b}}$) at different vertical positions}
        \label{fig_surface_res}
    \end{minipage}
\end{figure}

Since the reconstructed positions of surface events tend to be more inward than expected, resulting in a larger discrepancy compared to bulk events, the resolutions of bulk and surface events are estimated separately.
Bulk event resolution is estimated based on simulations. By using the patterns obtained from simulations as inputs for the reconstruction algorithms, the resolution of bulk events can be evaluated through the discrepancy between the reconstructed positions and the primary positions in MC. The primary positions are set to be within 400 mm from the center of TPC. 
As illustrated in figure~\ref{fig_bulk_res}, there is a significant enhancement in the bulk resolution of both the PAF and TM methods comparing to the COG method. Within this comparison, the PAF method provides a better performance over TM. Moreover, PAF method depends on physical model, which is more likely to describe the real physical process. Under the circumstances of high photon signal noises, PAF can more accurately reflect the spatial distribution of the photon signal, thereby avoiding the trap of local optima. Therefore, the position-related analysis are carried out using PAF method, while TM provides a supplementary verification. For clarity, in the following discussion, only the results obtained by the PAF method will be presented.

The position of surface events are expected to be reconstructed at the edge of the TPC.
$^{210}\textrm{Po}$ events, characterized by an alpha decay with energy of 5.3 MeV, are utilized to assess the surface resolution by evaluating the standard deviation of their radial distributions.
The ionization electron clouds of surface events may touch the reflector during the drift process, causing some of the ionization electrons to be adsorbed. This results in a suppressed $S2$ signal. Utilizing this feature, resolution in terms of different $S2$ charge can be evaluated, as shown in figure~\ref{fig_surface_res}. 
As mentioned before in Sec.\ref{subsec_cor}, the surface resolution is estimated based on $S2$ bottom charge.
From the five curves representing different vertical positions, it can be observed that the closer to the liquid surface, the better the reconstructed resolution. For signals with a small charge, the light pattern tends to be more diffused, making accurate reconstruction more difficult. Conversely, for signals with larger charges, the resolution degrades as the PMTs may become saturated.

\subsection{Uniformity}
\label{subsec_uniformity}
\begin{figure}[!b]
    \centering
    \includegraphics[width=\linewidth]{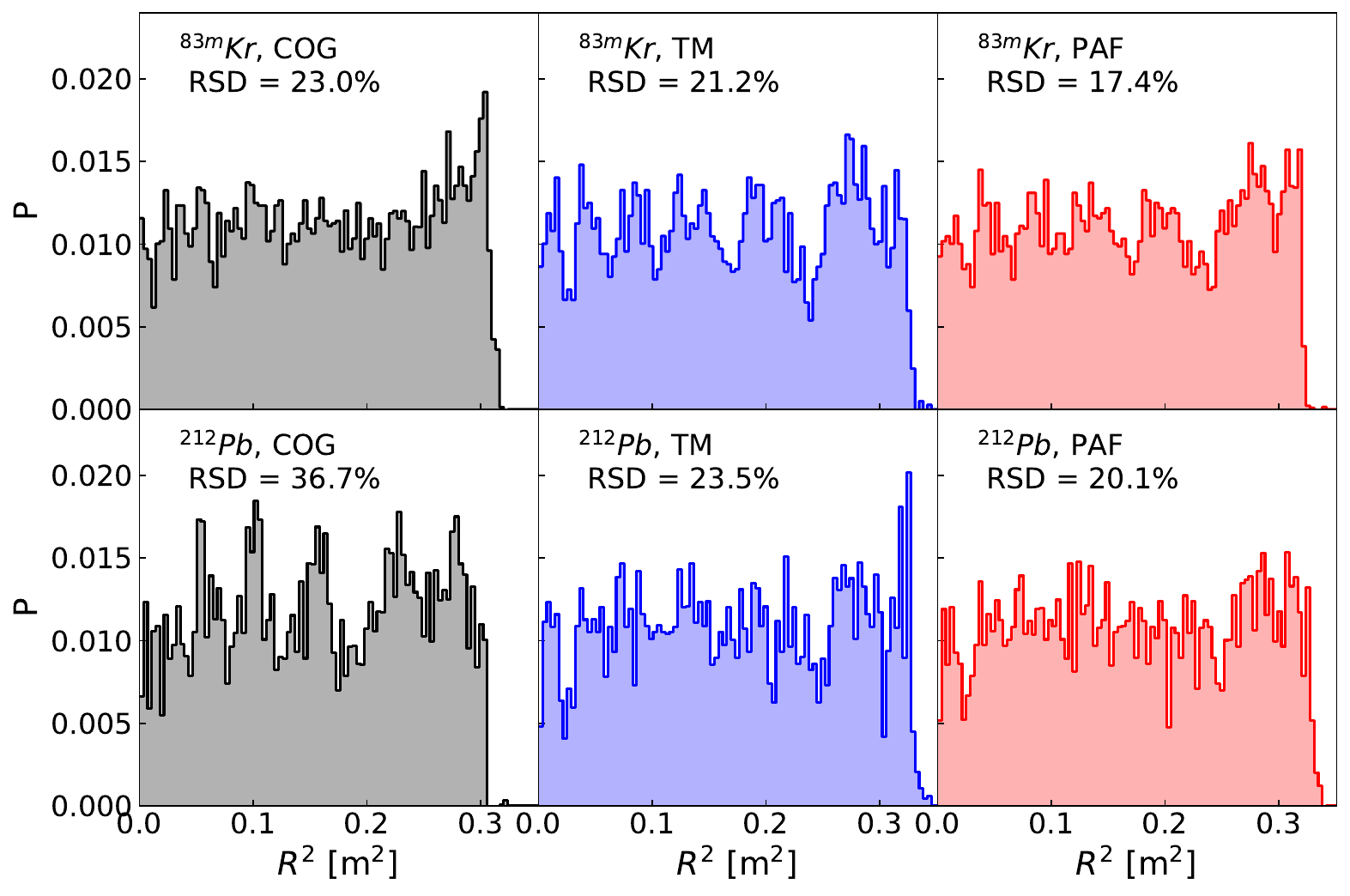}
    \caption{$R^{2}$ distribution of $^{83m}$Kr (top) and $^{212}$Pb (bottom) using different type of reconstruction methods.}
    \label{fig_rsd}
\end{figure}
During data processing, calibrations are done to help us understand the response of the detector. Calibration sources such as $^{83m}$Kr, $^{220}$Rn are supposed to be uniformly-distributed. Thus, the normalized spatial distribution $P(r^2)$ of these events can be used to evaluate the performance of position reconstruction.
The uniformity can be quantified by relative standard deviation (RSD) after binning, which is defined as
\begin{equation}\label{RSD}
\begin{aligned}
    \text{RSD} =\frac{\sqrt{\overline{P(n)^2}-[\overline{P(n)}]^2}}{\overline{P(n)}}, \text{where}\ \
    \overline{P(n)}=\frac{\sum\limits_{n=0}^{n_{\rm crit}}P(n)}{n_{\rm crit}},\ \ \overline{P(n)^2} = \frac{\sum\limits_{n=0}^{n_{\rm crit}}P(n)^2}{n_{\rm crit}},
\end{aligned}
\end{equation}
$n$ is the bin index, and $n_{\rm crit}$ corresponds to the bin where the normalized distribution falls to 20\% of its maximum value ($P(n_{\rm crit})=0.2\cdot P_{\rm max}$). Figure~\ref{fig_rsd} shows the radial distribution of $^{83m}$Kr, $^{220}$Rn using different reconstruction methods. 
Along the $R^{2}$ axis, from 0 to $0.36\ \rm m^{2}$, the range is divided into 100 bins. To ensure comparability between different methods, we averaged and standardized that $n_{\rm crit} = 90$. Correspondingly, the RSD quantifies the uniformity using data within $R^{2}<0.324\ \rm m^{2}$. The PAF method has the minimum RSD among all the three methods, indicating its best performance.

\subsection{Robustness}
\begin{figure}[!b]
    \centering
    \begin{subfigure}{0.45\textwidth}
        \includegraphics[width=\textwidth]{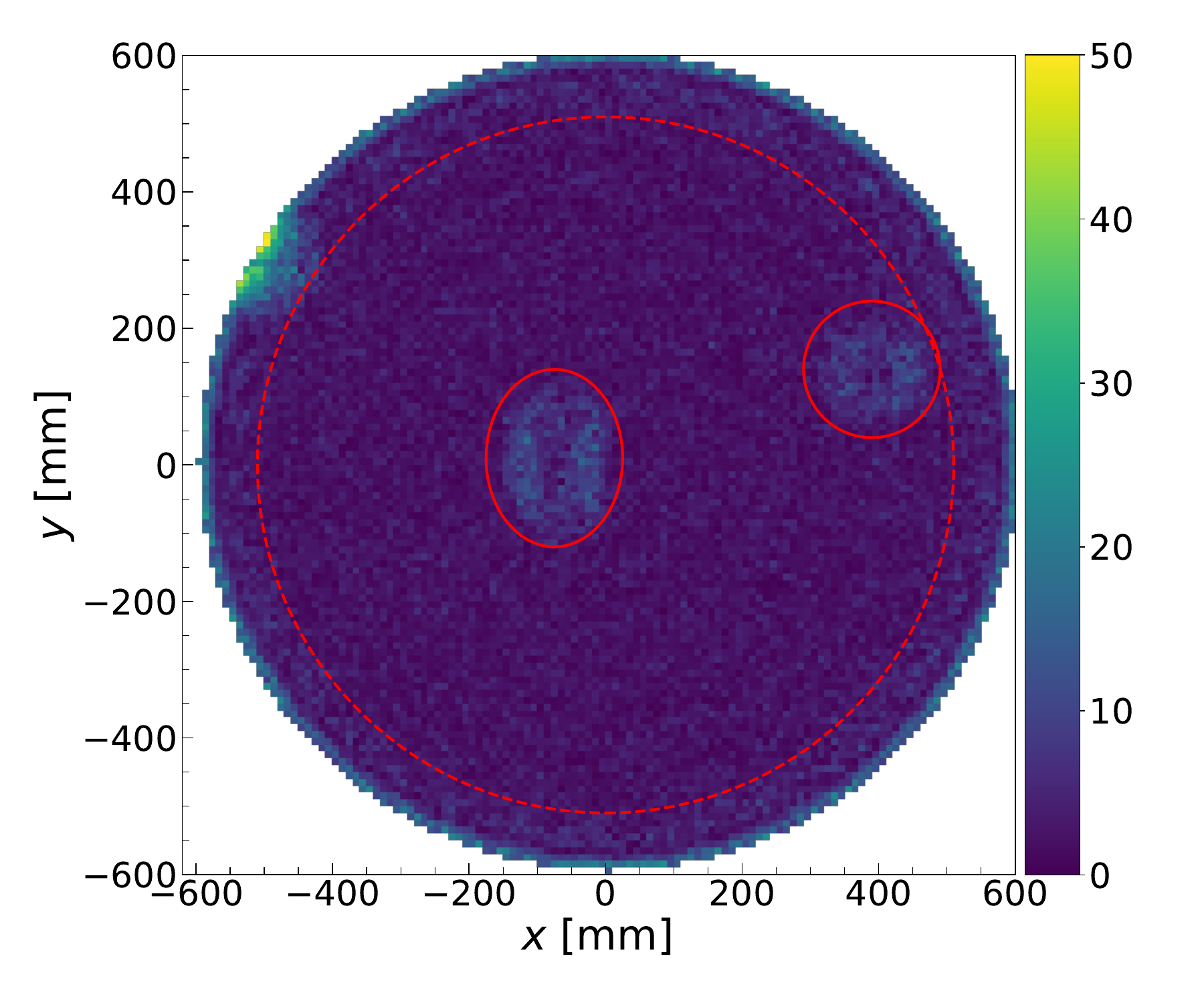}
        \caption{Run0}
        \label{fig:robustness_Run0}
    \end{subfigure}
    \begin{subfigure}{0.45\textwidth}
        \includegraphics[width=\textwidth]{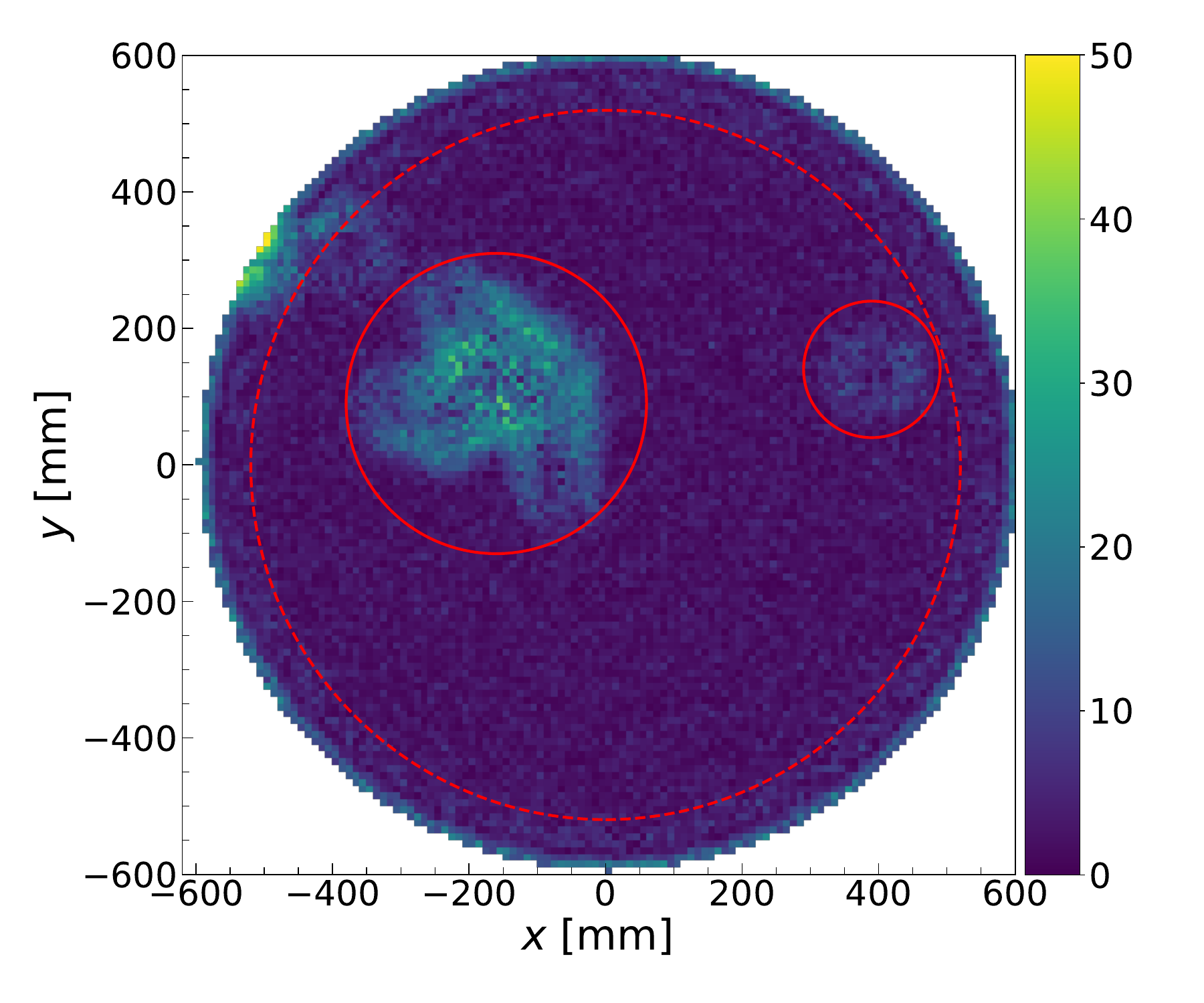}
        \caption{Run1}
    \label{fig:robustness_Run1}
    \end{subfigure}
    \caption{Robustness assessment using MC at $Q_{S2_{\rm t}}$ = 6000 PE. The color indicates the deviation at each position. Solid lines represent the off-PMT regions and dashed lines denote the FV boundaries.}
    \label{fig:robustness}
\end{figure}
\begin{figure}[h]
    \centering
    \includegraphics[width=0.55\linewidth]{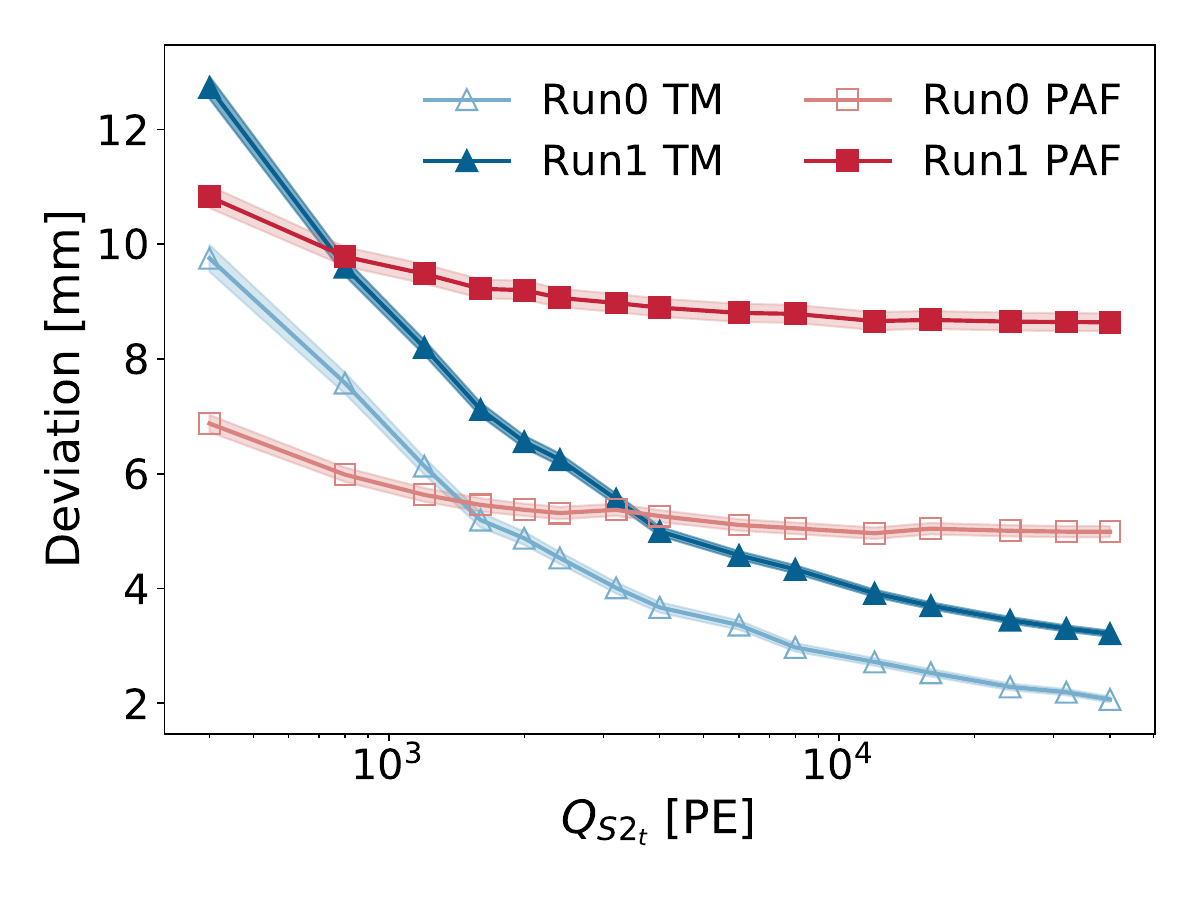}
    \caption{The dependence of deviation on charge in off-PMT regions (The colored bands represent the error bars).}
    \label{fig_dev_offPMT}
\end{figure}

Robustness in position reconstruction refers to the accuracy and reliability of the reconstruction despite missing or degraded data, such as when certain PMTs fail to collect light signals. It quantifies the reconstruction deviation under these compromised conditions. During the data-taking process, four top PMTs in Run0 and additional eleven PMTs in Run1 were turned off individually (figure~\ref{fig_RMT_map}), presenting a challenge for position reconstruction. Robustness is assessed using MC simulations by uniformly scattering points in the $x-y$ plane and comparing primary and reconstructed positions. The saturation effect of the top PMTs is not considered in simulation. Figure~\ref{fig:robustness} shows the deviations at $Q_{S2_{\rm t}}$ = 6000 PE using PAF method under the running conditions of Run0 and Run1. 
It can be obviously observed that the reconstructed positions show greater deviations around the off-PMT regions (red solid lines in figure~\ref{fig:robustness}). To further clarify the robustness, we calculate the average deviation within the off-PMT regions. Figure~\ref{fig_dev_offPMT} shows the dependence of robustness on charge, with the PAF method outperforming the TM method in the low-energy region. For a typical $S2_{\rm t}$ signal at 6000 PE, the mean deviations within the off-PMT regions are 5.1 (3.4) mm and 8.8 (4.6) mm for Run0 and Run1 using PAF (TM) method, respectively. Compared to the situation where the PMTs are in normal operation status (figure~\ref{fig_bulk_res}), the deviation worsens by 1 to 8 mm.

\section{Surface background model}
\label{sec_surface_model}
To investigate the contribution of surface background events, a surface background model is required to predict the distribution along $R$ and within the parameter space of $\log_{10}(Q_{S2_{\rm b}}/Q_{S1})$ vs. $Q_{S1}$. This parameter space is used for analyzing WIMP candidates~\cite{pandax2024}.
The model is based on the position reconstruction algorithms discussed in the previous sections. As mentioned in Sec.~\ref{subsec_resolutions}, the $R$ distribution of $^{210}$Pb can be modeled by studying the $S2$ characteristics of $^{210}$Po alpha events, ultimately estimating the number of surface background events within energy ROI and FV. The process of building the surface background model involves the following steps: creating probability density functions (PDFs), 2-D slicing along $z$ and $Q_{S2_{\rm{b}}}$, using a re-weighting technique, fittings of model parameters, interpolating along $Q_{S2_{\rm{b}}}$, combination and normalization. 

\begin{figure}[htbp]
\centering
    \begin{minipage}[t]{\linewidth}
        \centering
        \includegraphics[width=0.6\linewidth]{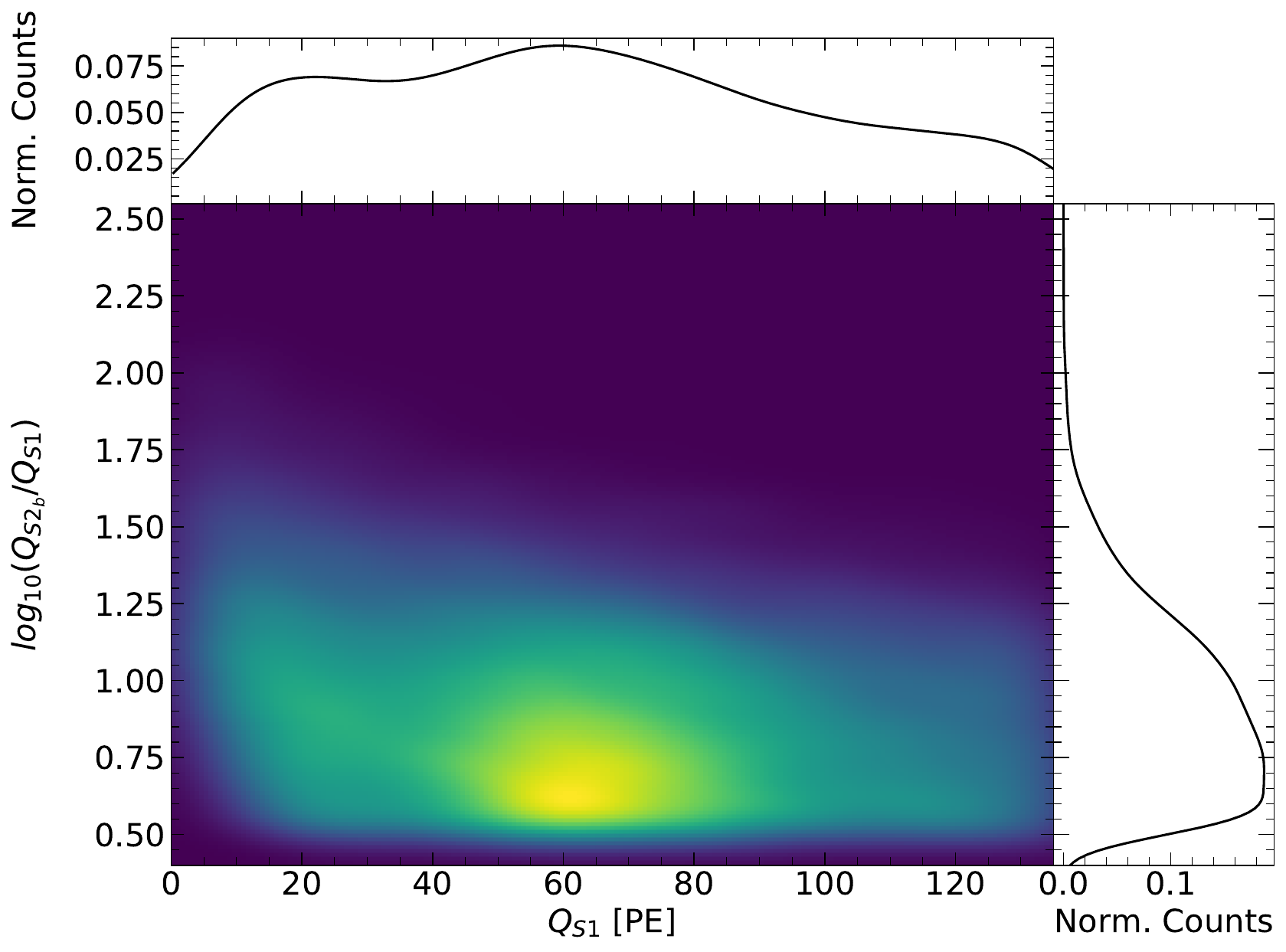}
        \caption{Smoothed distribution of the events outside the PTFE wall in $\log_{10}(Q_{S2_{\rm b}}/Q_{S1})$ vs. $Q_{S1}$}
        \label{fig_KDE}
    \end{minipage}
\end{figure}

\textbf{Creating PDFs} --- 
Assuming that the events in the energy ROI reconstructed outside the PTFE wall originate from the PTFE surface, then the smoothed three-dimensional distribution of $S1$, $S2$, and $z$ from these events can be used as the PDF of the model. Figure~\ref{fig_KDE} demonstrates the PDF and the bright spot comes from the L-shell Auger electrons and X-ray of $\rm{}^{210}Pb$ decay. 

\begin{figure}[!htbp]
    \centering
    \begin{subfigure}[t]{0.32\textwidth}
        \includegraphics[width=\textwidth]{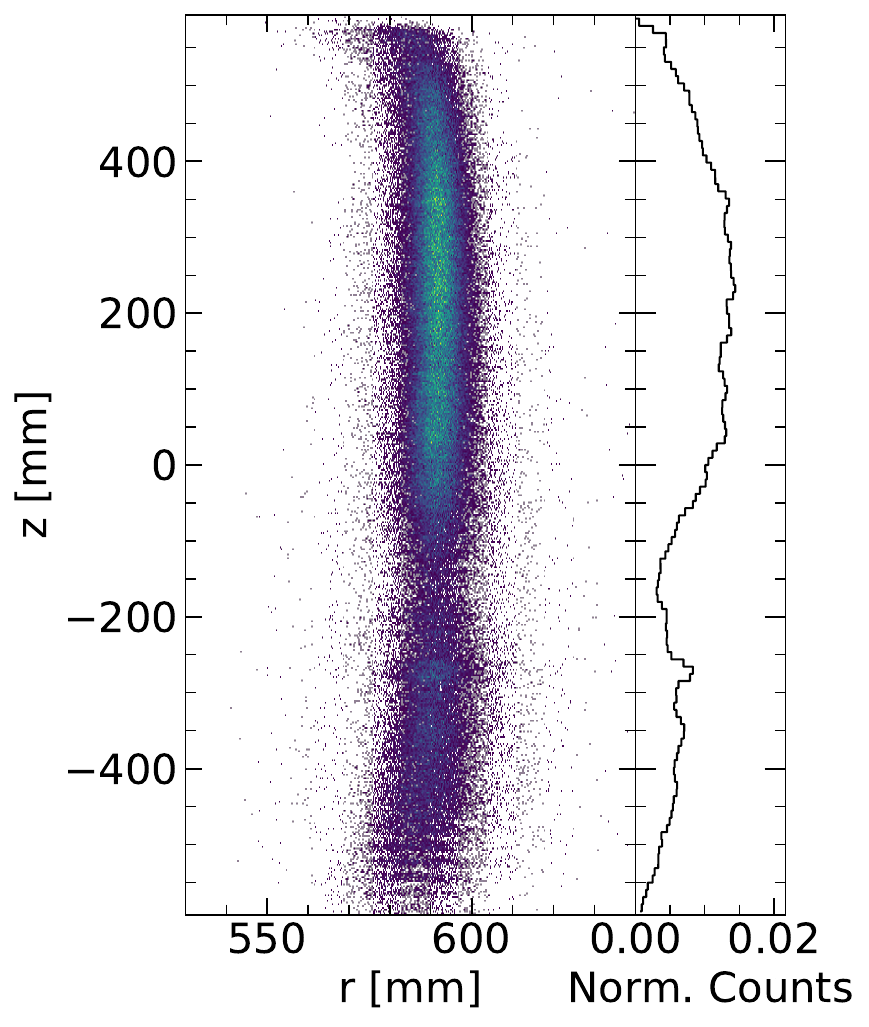}
        \caption{$R$ vs. $z$}
        \label{fig:Po_z}
    \end{subfigure}
    \begin{subfigure}[t]{0.32\textwidth}
        \includegraphics[width=\textwidth]{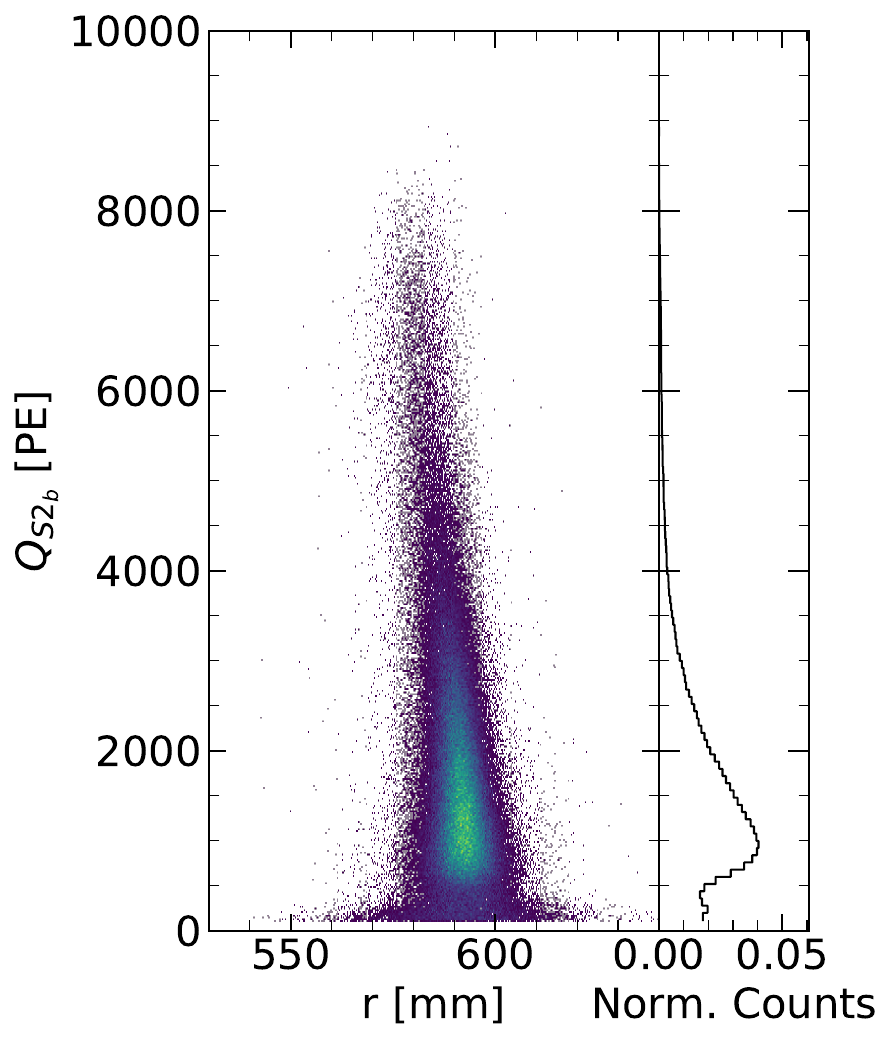}
        \caption{$R$ vs. $Q_{S2_{\rm b}}$}
    \label{fig:Po_s2}
    \end{subfigure}
    \begin{subfigure}[t]{0.32\textwidth}
        \includegraphics[width=\textwidth]{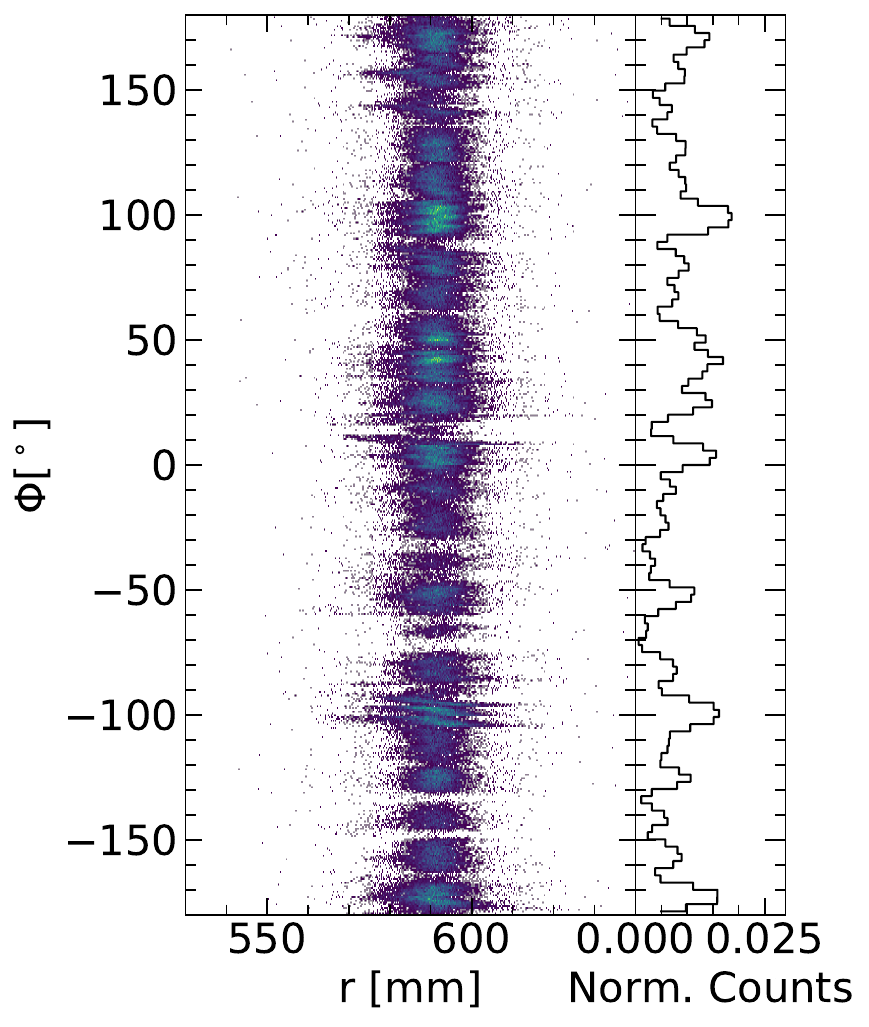}
        \caption{$R$ vs. $\rm{\Phi}$}
    \label{fig:Po_phi}
    \end{subfigure}
    \caption{Radial dependence of $^{210}$Po surface events on vertical position, $S2_{\rm{b}}$ charge and azimuth angle.}
    \label{fig:Po_r_dependence}
\end{figure}

\textbf{2-D slicing along $z$ and $Q_{S2_{\rm{b}}}$} ---
The radial distribution of surface events depends on $S2_{\rm{b}}$ charges and vertical position (figure~\ref{fig:Po_z} and figure~\ref{fig:Po_s2}). With GPC mentioned in Sec.~\ref{subsec_cor}, the radial dependence on $\Phi$ has been eliminated, but their PDFs are not uniform (figure~\ref{fig:Po_phi}). Therefore, a two-dimensional slicing of the $^{210}$Po samples in terms of $z$ and $S2$ is performed and the properties of each slice are studied in the $R-\Phi$ plane.

\textbf{Using a re-weighting technique} --- 
When selecting $^{210}$Po events as a template, an event whose $S1$ is accidentally paired with a random $S2$ should not be included. For small $S2$, this mis-pairing rate is high. Therefore, a lower threshold for $S2$ has been set to ensure a purer template. 
Consequently, the $S2$ distribution of the $^{210}$Po samples differs from that of low-energy surface background events in the ROI. Using a re-weighting technique, we can better model the distribution of low-energy surface events.

\begin{figure}[!htbp]
\centering
    \begin{minipage}[t]{0.48\linewidth}
        \centering
        \includegraphics[width=\linewidth]{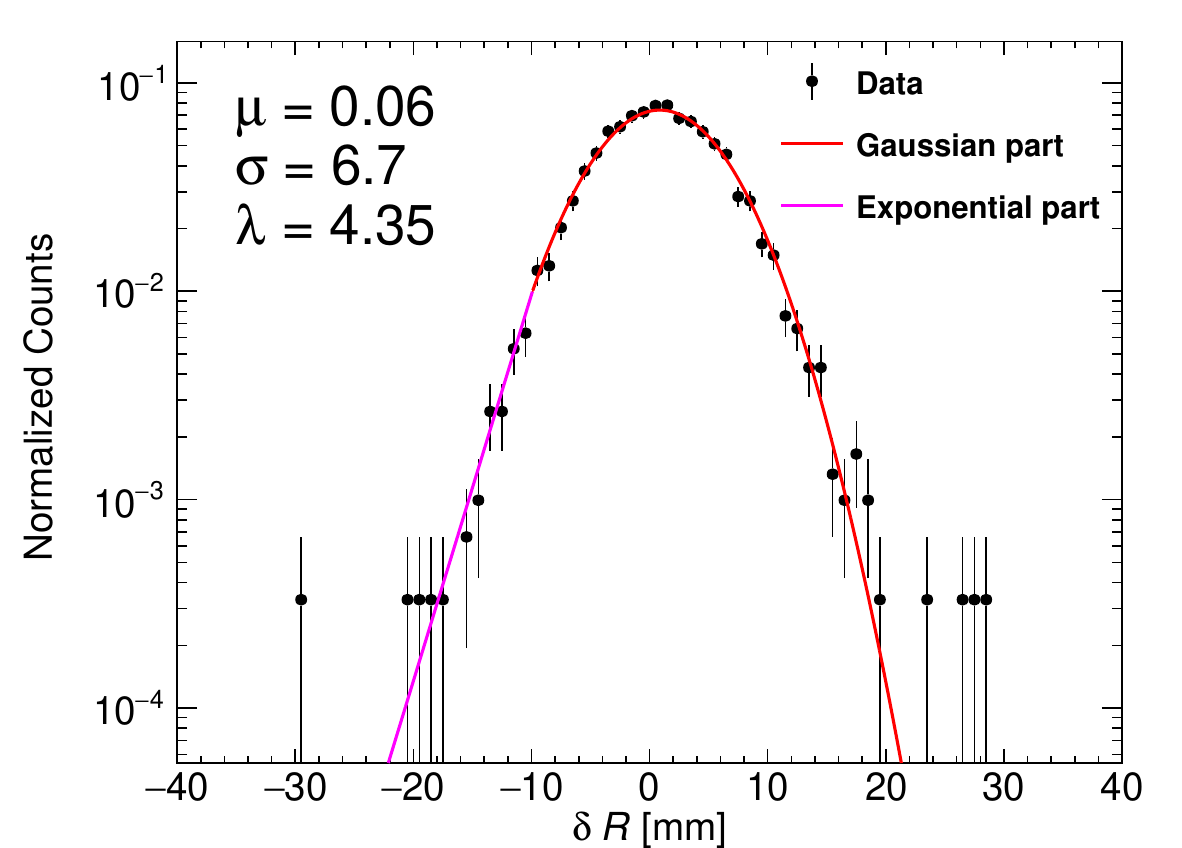}
        \caption{Reduced $R$ distribution relative to the median curve of $R-\Phi$ distribution. Exponential Gaussian tail function is used for fitting, which is composed of a Gaussian term and an exponential term.}
        \label{fig_deltaR}
    \end{minipage}
    \hfill
    \begin{minipage}[t]{0.48\linewidth}
        \centering
        \includegraphics[width=\linewidth]{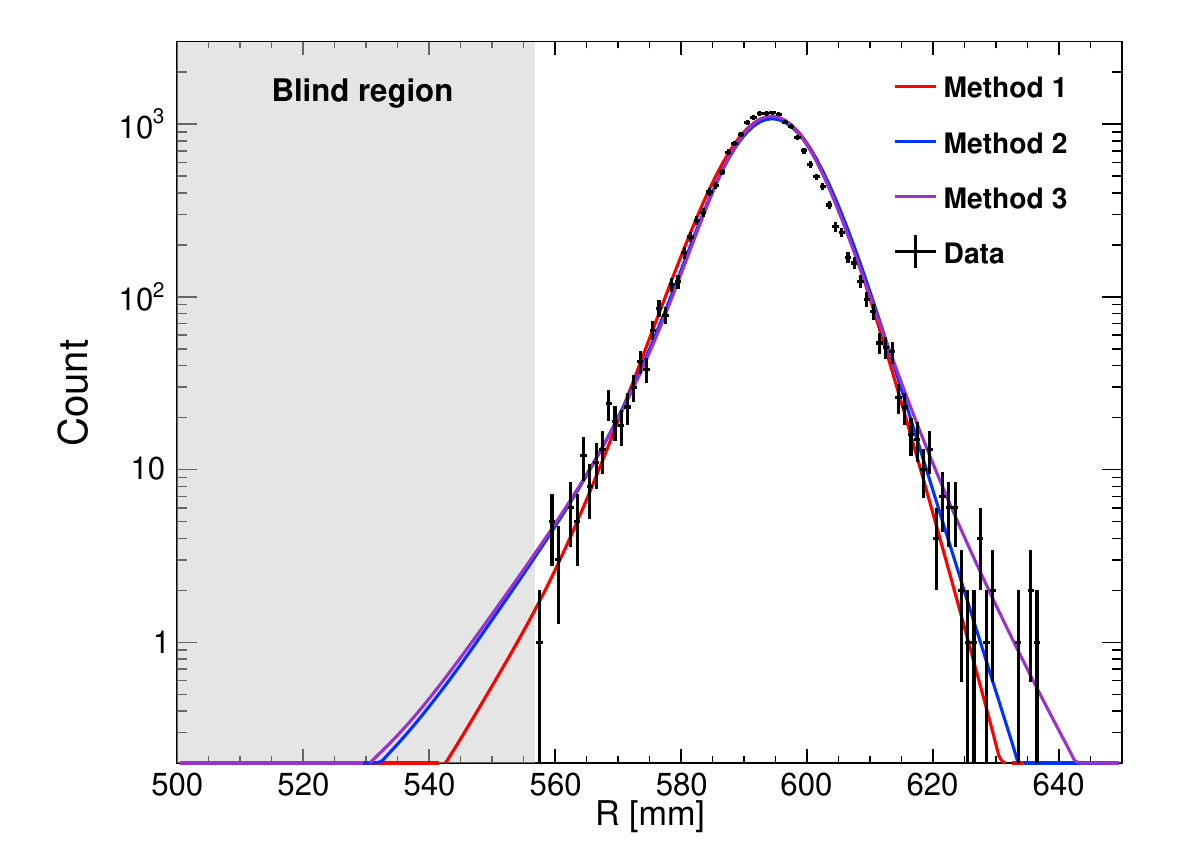}
        \caption{Radial distribution of surface background model using three fitting methods. The corresponding fitting functions are provided in table~\ref{tab_surface}.}
        \label{fig_surface_model}
    \end{minipage}
\end{figure}

\textbf{Fittings of model parameters} --- 
For each slice, set the median curve in the $R-\Phi$ plane as the zero point. Then, a reduced radial distribution ($\delta R$) relative to this median curve can be obtained (figure~\ref{fig_deltaR}). This distribution is Gaussian-like in shape, and can be fitted using various Gaussian-like functions (table~\ref{tab_surface}) to determine the model parameters such as the center value $\mu$ and the standard deviation $\sigma$. Figure~\ref{fig_deltaR} provides an example slice where the fitting function consists of a Gaussian term and an exponential term, allowing the  derivation of $\mu, \sigma$ and a parameter $\lambda$, which describing the decreasing event rate inwards along the radial direction.

\textbf{Interpolating along $\boldsymbol{Q_{S2_{\rm b}}}$} ---
By combining the fitting results of each ($z$, $Q_{S2_{\rm b}}$) slice, we can determine $S2$ charge-dependent parameters for different vertical positions. For example, figure~\ref{fig_surface_res} illustrates the case of the parameter $\sigma$. To build a smoother model, we further interpolate along the $Q_{S2_{\rm b}}$. Consequently, given any pair of values ($z$, $Q_{S2_{\rm b}}$), their radial distribution can be determined.

\textbf{Combination and normalization} ---
For WIMP analysis, only events below the NR median are considered relevant. By combining all slices within ROI from the PDFs (figure~\ref{fig_KDE}) mentioned before, the shape of radial distribution for surface events is established. Normalizations are done according to the low energy events reconstructed outside the PTFE wall. Finally, figure~\ref{fig_surface_model} shows the radial distribution of surface background model using different fitting methods provided in table~\ref{tab_surface}. For a blind analysis, we mask the data in the blind region where $R^2 < 0.31\ \rm{m}^2$. The consistency between the model and data can be evaluated by the $\chi^2/\rm NDF$ value in the region between the blind boundary and the peak (table~\ref{tab_surface}). The models effectively describe the data outside the blind region, enabling estimation of the surface background contribution inside the blind region.
\begin{table}[htbp] 
\centering
\tiny
  \begin{tabular}{|c|c|cc|cc|}
    \hline
    \multirow{2}{*}{Method} & \multirow{2}{*}{Function} & \multicolumn{2}{|c|}{Run0} & \multicolumn{2}{|c|}{Run1} \\ 
      & & count & $\chi^2/\rm NDF$ & count & $\chi^2/\rm NDF$ \\ \hline
     1& \multicolumn{1}{l|}{
$
f(x) = 
\begin{cases} 
A_{1} \exp \left( - \frac{(x - \mu)^2}{2 \sigma_1^2} \right) & \text{if } x < \mu \\
A_{2} \exp \left( - \frac{(x - \mu)^2}{2 \sigma_2^2} \right) & \text{if } x \geq \mu
\end{cases}
$
    }& 0.03  & 2.3              & 0.05  & 4.4              \\
     2&\multicolumn{1}{l|}{
$
f(x) = 
\begin{cases} 
A_{1}\exp \left( - \frac{(x - \mu)}{\lambda} \right) & \text{if } x < \mu - \sigma \\
A_{2}\exp \left( - \frac{(x - \mu)^2}{2 \sigma^2} \right) & \text{if } x \geq \mu + \sigma
\end{cases}
$     
    }& 0.12  & 2.9              & 0.26  & 2.3              \\
     3&\multicolumn{1}{l|}{
$
f(x) = 
\begin{cases} 
A_{1} \exp \left( - \frac{(\mu - x)}{\lambda_1} \right) & \text{if } x < \mu - \sigma \\
A_{2} \exp \left( - \frac{(x - \mu)^2}{2 \sigma^2} \right) & \text{if } \mu - \sigma \leq x \leq \mu + \sigma \\
A_{3} \exp \left( - \frac{(x - \mu)}{\lambda_2} \right) & \text{if } x > \mu + \sigma
\end{cases}
$
    }& 0.13  & 2.3              & 0.19  & 2.5              \\ \hline\hline
     \multicolumn{2}{|c|}{Average} &\multicolumn{2}{c|}{$0.09 \pm 0.06$} &\multicolumn{2}{c|}{$0.17 \pm 0.11$} \\ \hline
  \end{tabular}
  \caption{Surface background estimation.}
  \label{tab_surface}
\end{table}

\begin{figure}[!h]
\centering
    \begin{minipage}[t]{\linewidth}
        \centering
        \includegraphics[width=0.6\linewidth]{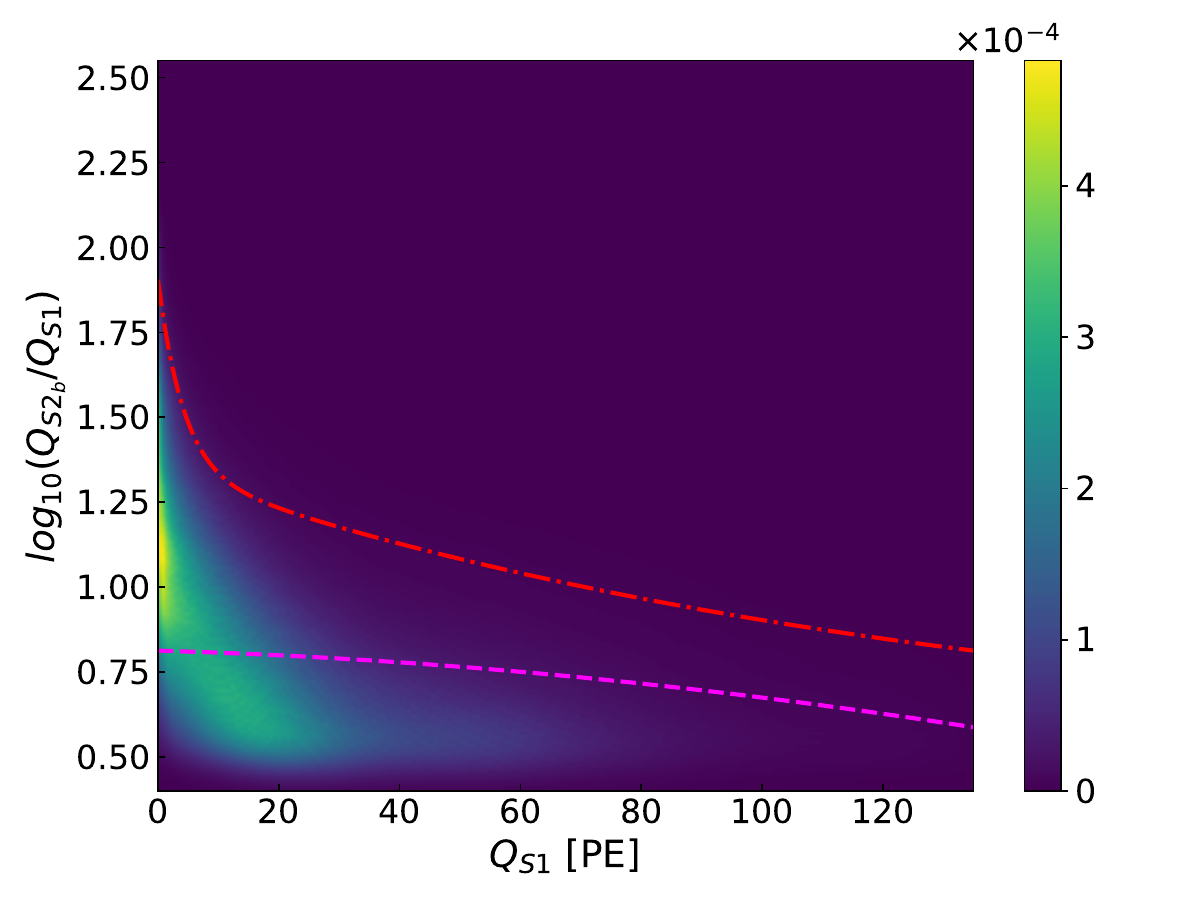}
        \caption{Distribution of surface background events inside FV in $\log_{10}(Q_{S2_{\rm b}}/Q_{S1})$ vs. $Q_{S1}$, overlaid with NR median curve (red line) and 99.5\% NR acceptance boundary (magenta line)}
        \label{fig_surface_PDF_in_FV}
    \end{minipage}
\end{figure}
Based on the surface background model and other background sources, FV is optimized utilizing a figure-of-merit~\cite{Cowan2012_DiscoverySF}. The $R^2$ boundary is determined to be $0.26 \rm{}\ m^2$ and $0.27 \rm{}\ m^2$~\cite{pandax2024} for Run0 and Run1 by the dominance of surface background events. The number of surface background events can then be estimated by integrating radially from 0 to the $R^2$ boundary. Table~\ref{tab_surface} summarizes the estimation of surface events using three fitting functions. The final result is the average of the three fitting functions, with the error represented by their standard deviations. Figure~\ref{fig_surface_PDF_in_FV} shows the distribution of surface background events inside FV in the parameter space of $\log_{10}(Q_{S2_{\rm b}}/Q_{S1})$ vs. $Q_{S1}$. For WIMP analysis, the energy ROI is defined as $Q_{S1}\in (2, 135)\rm\ PE$, below NR median curve and above 99.5\% NR acceptance boundary. Within the optimized FV and energy ROI, the surface background is estimated to be $0.09 \pm 0.06$ events for Run0 and $0.17 \pm 0.11$ events for Run1, respectively~\cite{pandax2024}.

\section{Summary}
\label{sec_summary}
In summary, we present two position reconstruction algorithms developed for the PandaX-4T experiment. Tailored to the specific characteristics of the PandaX-4T detector, these algorithms include advanced corrections to further enhance reconstruction resolution. After evaluating both methods based on their resolution and uniformity performance, the PAF method was chosen for primary position reconstruction, while the TM method served as a verification tool. For a typical $S2$ signal with $Q_{S2_{\rm b}}$ = 1500 PE (equivalently, $Q_{S2_{\rm t}}$ = 6000 PE), the bulk and surface event resolution reaches approximately 1.0 mm and 4.4 mm. The uniformity is estimated using $^{83m}$Kr and ${^{212}}$Pb events, resulting in an RSD of around 20\%. The Robustness study indicates that the average deviations are 5.1 mm and 8.8 mm for Run0 and Run1 within off-PMT regions when $Q_{S2_{\rm t}}$ = 6000 PE. 
The PAF method also forms the basis for a data-driven surface background model, which estimates the surface background contribution as $0.09 \pm 0.06$ events for Run0 and $0.17 \pm 0.11$ events for Run1. Beyond enhancing the immediate accuracy of position reconstruction, these improvements contribute significantly to optimizing energy reconstruction, correcting detector non-uniformity, and boosting the signal-to-noise ratio. Looking ahead, these advancements will not only refine background modeling but also improve the overall sensitivity of the experiment, paving the way for future breakthroughs in dark matter detection.

\section*{Acknowledgment}
This work has been supported by the Ministry of Science and Technology of China (No. 2023YFA1606204), the National Natural Science Foundation of China (No. 12222505, No. 12205181, No. 12325505). We also thank the sponsorship from the Chinese Academy of Sciences Center for Excellence in Particle Physics (CCEPP), Hongwen Foundation in Hong Kong, New Cornerstone Science Foundation, Tencent Foundation in China, and Yangyang Development Fund.

%% If you have bib database file and want bibtex to generate the
%% bibitems, please use
%%
%%  \bibliographystyle{elsarticle-num} 
%%  \bibliography{<your bibdatabase>}

%% else use the following coding to input the bibitems directly in the
%% TeX file.

%% Refer following link for more details about bibliography and citations.
%% https://en.wikibooks.org/wiki/LaTeX/Bibliography_Management

\bibliographystyle{elsarticle-num}
\bibliography{biblio}
\end{document}